\documentstyle[12pt,epsf]{article}
\def\appendix#1{
  \addtocounter{section}{1}
  \setcounter{equation}{0}
  \renewcommand{\thesection}{\Alph{section}}
 \section*{Appendix \thesection\protect\indent \parbox[t]{11.715cm} {#1}}
  \addcontentsline{toc}{section}{Appendix \thesection\ \ \ #1}
  }

\newcommand{\newsection}{
\setcounter{equation}{0}
\section}

\def\bea{\begin{eqnarray}}
\def\eea{\end{eqnarray}}
\def\be{\begin{equation}}
\def\ee{\end{equation}}
\newcommand{\tr}[1]{\:{\rm tr}\,#1}

\def\e{{\,\rm e}\,}
\def\d{\partial}
\def\D{\delta}

\def\la{\longrightarrow}
\def\lp{\mapsto}
\def\us{U(N_1)\times\ldots\times U(N_m)}

\newcommand{\rf}[1]{(\ref{#1})}

\newcommand{\Ker}{\mathop{\rm Ker}}
\renewcommand{\Im}{\mathop{\rm Im}}
\begin{document}

\begin{flushright}
ITEP--TH--7/99
\end{flushright}
\vspace{1.5cm}

\begin{center}
{{\huge {\bf Solitons on Branes}
}}\\
\vskip 0.3truein
{\bf G. W. Semenoff}\footnote{Research supported by NSERC of Canada and
the Niels Bohr Fund of Denmark and NATO Grant CRG 970561.  
Permanent address: {\it Department of Physics
and Astronomy, University of British Columbia, 6224 Agricultural Road, 
Vancouver, British Columbia, Canada V6T 1Z1}. E-mail: semenoff@alf.nbi.dk}
and {\bf K. Zarembo}\footnote{Research supported by a NATO Science
Fellowship, NSERC of Canada, INTAS grant 96-0524, RFFI grant 98-01-00327.
Permanent address: {\it Department of Physics
and Astronomy, University of British Columbia, 6224 Agricultural Road, 
Vancouver, British Columbia, Canada V6T 1Z1 } and {\it Institute for 
Theoretical and Experimental Physics, B. Cheremushkinskaya 25, 117259, Moscow,
Russia}. E-mail: zarembo@alf.nbi.dk}
\vskip 0.3truein
{\it The Niels Bohr Institute\\
Blegdamsvej 17\\
DK-2100 Copenhagen 0\\
Denmark}

\vskip 1.0truein
{\bf Abstract}
\vskip 0.3truein
\end{center}
We examine the possibility that gauge field configurations on stacks
of parallel Dp-branes support topological solitons.  We give an exhaustive
list of possible soliton charges for $p\leq 6$.  
We also discuss how configurations carrying the soliton charges 
can be constructed from intersecting branes.

\newsection{Introduction}

One of the most intriguing features of D-branes is that they provide the
possibility of embedding super-symmetric gauge theories in various
dimensions into string theory \cite{wit95} (see \cite{gk98} for a review).
In many cases, these field theories possess soliton states which are 
associated
with topologically non-trivial configurations of the fields. 
The fact that the gauge fields on D-branes can form such 
topologically non-trivial configurations appears to be important
in the interpretation of D-brane charges  \cite{mm97,wit98,gar98,guk99,sha99} 
and potentially could lead to a deeper understanding of the origin of the 
D-branes themselves \cite{sen98,wit98,hor98,yi99,bgh99,ahh99}. 

The topological charge on the world-volume of a D-brane couples
to the space-time Ramond-Ramond fields of type II superstring theory via 
Chern-Simons terms in the world-volume action \cite{dou95,ghm96,cy97}. 
This allows an alternative interpretation of the topological charge 
as the RR charge due to the presence of 
lower-dimensional branes which are either suspended between the higher 
dimensional branes \cite{str95} or   
embedded in the world-volume of the higher dimensional branes \cite{dou95}.
An example is the description of BPS magnetic monopoles in $d=4$, 
${\cal N}=4$ super-Yang-Mills theory which can be viewed either as 
conventional BPS
monopoles of the Coulomb branch of the gauge theory 
or as D-strings stretched between parallel D3-branes 
\cite{str95,dia96}.  Also, instantons in that theory can be described 
alternatively as either
conventional Yang-Mills instantons or (the T-dual of) D0-branes 
immersed in the world-volume of D4-branes \cite{dou95,vaf95,dou96,bp97,tay98}.
Our aim in this Paper is to give a complete topological classification of
world-volume defects and their string interpretation for all stable BPS 
Dp-brane states in type IIA and IIB string theories, with $p\leq 6$ and 
which preserve half of the ten-dimensional ${\cal N}=2$ super-symmetry.

The most general configurations of this type contain
$m$ parallel Dp-branes each carrying RR charges $N_1,\ldots,N_m$.
In the limit when their positions coincide, 
the internal symmetry group of such a configuration 
is $G=U(N)$ \cite{wit95}, where $N=\sum N_i$ is the
total RR charge. When the branes are separated to form parallel stacks of 
$N_1,N_2,\ldots$ coinciding branes, this 
symmetry is spontaneously broken to $$H=U(N_1)\times\ldots\times U(N_m)~~~.$$
This situation generically
allows for the existence of topological defects \cite{sch89}.
The soliton states arise because the space of classical vacua is a 
manifold with non-trivial topology.  The vacua, which are field configurations
of the supersymmetric gauge theory 
with zero potential energy, are flat connection  gauge fields residing in the 
unbroken subgroup of the gauge group and values of scalar 
fields which minimise the potential. Finiteness of the 
energy imposes boundary conditions on any legitimate field configuration,
requiring it to take vacuum values at infinity. The possible boundary
conditions are classified by topologically distinct mappings from the
sphere at infinity to the space of vacua.  This leads to the existence
of topologically distinct sectors in the space of all field configurations. 
The charges which distinguish these
sectors take values in the $\pi_{p-1}$ homotopy group
of the vacuum manifold.

The instanton charges are related to topologically non-trivial pure gauge
fields of the unbroken subgroup and
are classified by
$$\pi_{p-1}(\us)~~~.$$ 
Another type of topological defect, called a topological soliton, 
is classified by mappings from the  sphere at infinity   
to the space of equilibrium values of internal scalar fields.
If the positions of branes are held fixed, the classical vacua  are
parametrised by the coset space $U(N)/U(N_1)\times\ldots\times U(N_m)$
and the soliton states are classified by the homotopy groups
$$\pi_{p-1}(U(N)/U(N_1)\times\ldots\times U(N_m))$$ 
We remind the reader that topological arguments alone do not guarantee
the existence of stable solutions of the classical equations of motion
in a given topological sector.

The homotopy groups of a coset space $G/H$ occur in exact sequences of the 
fibration $G\rightarrow G/H$:
\be\label{es}
\ldots\la\pi_{p-1}(G)\stackrel{f_1}{\la}\pi_{p-1}(G/H)
\stackrel{f_2}{\la}\pi_{p-2}(H)\la\ldots.
\ee
In all of the cases  which we shall consider, the exactness of this sequence 
unambiguously fixes the necessary homotopy groups. Also, certain information
about the nature of the resulting 
topological solitons can be obtained from the structure of
the homomorphisms $f_1$ and $f_2$.

If $f_1$ is an isomorphism,
the corresponding topological defects are instantons of the broken
gauge group \cite{cho79}. Indeed, 
the homomorphism $f_1$ is induced by the
projection map $G\rightarrow G/H$. An explicit realization of the 
projection map is $$U\lp U^{-1}\phi^i U~~~~,$$  where $\phi^i$ is a set of 
diagonal $N\times N$ matrices. 
Interpreting the eigenvalues of $\phi^i$, $i=1\ldots 9-p$ as transverse 
coordinates of $N$ D-branes 
 and denoting by $\Phi^i$  
corresponding scalar fields in the world-volume theory, we find that
a field configuration going to the pure gauge at 
infinity:
$$
\Phi^i\la  U^{-1}\phi^i U,~~~~A\la U^{-1}d U,
$$
carries a topological charge taking values in $\pi_{p-1}(G/H)$. 
The homotopy class of this configuration coincides with the 
homotopy class of the gauge transformation $U$ at infinity 
which takes values in $\pi_{p-1}(G)$. 

For $N$ sufficiently large, the homotopy 
groups responsible for instantons are periodic in $p$ \cite{dnf}:
\be\label{bott}
\pi_1(U(N))=\pi_3(U(N))=\pi_5(U(N))=Z.
\ee
Thus, topological arguments suggest that there are instantons on 
even-dimensional branes.
There are following exceptions for low $N$:
\be
\pi_k(U(1))=0~(k>1),~~~~\pi_4(U(2))=\pi_5(U(2))=Z_2.
\ee
The exceptional case of $N=2$, $p=5$ is
discussed in Sec.~\ref{u2}.  

On the string side, instantons are described by D0-branes
embedded inside Dp-branes \cite{dou95,vaf95,dou96,tay97,bp97,tay98}.
As was mentioned in \cite{wit98},
the Bott periodicity \rf{bott} of the homotopy groups $\pi_n(U(N))$ 
appears to be encoded in the string description, since 
0-branes exist in type IIA string theory and can only couple to 
even-dimensional branes. 

If $f_2$ is an isomorphism, so that
$$
\pi_{p-1}(G/H)=\pi_{p-2}(H)~~~,
$$
or, more generally, if $f_2$ is an onto mapping, so that
$$
\pi_{p-1}(G/H)=\pi_{p-2}(H)/\pi_{p-2}(G)~~~,
$$ 
the corresponding defects
are generalised monopoles, as explained in Sec.~\ref{general}, and the
generic case is described by open D1-branes stretched between adjacent
higher-dimensional branes. Again, the D-brane picture matches the 
topological consideration, since, on the string side, 
we expect to find monopoles on
odd-dimensional branes of type IIB theory. On the other hand, 
the stable homotopy
groups responsible for monopoles are
\be\label{main}
\pi_{p-1}(U(N)/U(N_1)\times\ldots\times U(N_m))=
\left\{
\begin{array}{ll}
Z^{m-1}, & p {\rm ~odd} \\
0, & p {\rm ~even}
\end{array}
\right. .
\ee
For completeness, we sketch the calculation of these groups in  
Appendix~A. 

\newsection{Monopoles on D-branes}
\label{general}

The potential of the super-Yang-Mill's theory has the form
$$
V\propto {\rm Tr}\left(\sum_i(D\Phi^i)^2
-\sum_{i<j}\left[ \Phi^i,\Phi^j\right]^2\right) 
$$
and is minimised by classical fields $\Phi^i$ $i=1,...,9-p$
which commute with each 
other and are therefore simultaneously diagonalisable. 
Their distinct eigenvalues, $\phi^i_a$, $a=1,...,m$, each with
degeneracy $N_1,..,N_m$ are the positions of the $m$ stacks of parallel 
$Dp$-branes.  
In a field configuration with a topological defect, the scalar fields
must 
minimise the potential at infinite distances from the location of the defect 
and deviate from the minimum inside the core.
Even though the eigenvalues of $\Phi^i$ must therefore 
approach constants at infinity,
generally, the scalar fields themselves approach non-diagonal matrices
$\Phi^i= U^{-1}\phi^iU$.  
Furthermore, to minimise the classical energy, 
these matrices must be covariant 
constants.   The gauge fields therefore 
approach pure gauges $A\rightarrow U^{-1}dU$.

Generally, $U$ is not defined globally on the sphere
at infinity, $S^{p-1}$, but must have a 
singularity at some point on $S^{p-1}$.
In this case, the singular
gauge transformation which passes to the unitary gauge by removing 
$U$ and leaving $\Phi^i$ diagonal creates a Dirac string.  In this gauge, 
both the Dirac string and the gauge field
at infinity carry magnetic fluxes in the unbroken subgroup $H$.
This is a result of the fact that, 
from the definition of the boundary homomorphism $f_2$ \cite{dnf},
the singularity of $U$ resides in the unbroken subgroup $H$.  
This defines 
a mapping $V$ from the sphere $S^{p-2}$ which links the
intersection of the Dirac string with $S^{p-1}$ to 
$H$.  If the homotopy class of this mapping is represented by
an integral over $S^{p-2}$ 
of the $(p-2)$-form $\tr\omega^{p-2}$, where 
$\omega=V^{-1}dV$, then it is equal to the magnetic flux
through the sphere at infinity, $\int_{S^{p-1}} \tr F^{(p-1)/2}$, 
where $F$ is 
the field strength in the unbroken subgroup and the gauge field 
approaches $\omega$ on $S^{p-2}$.
The structure of the mapping $f_2$ between homotopy 
groups is such that, when it is an isomorphism, 
the magnetic flux in the unitary gauge is equal to the 
topological charge of the defect.  The corresponding topological
defects are generalised magnetic monopoles.
If $f_2$ is not an isomorphism, the topological charge is equal 
to the magnetic
flux modulo an instanton number of the gauge field $A$. 

These generalised monopoles are described by D-strings stretched between
stacks of Dp-branes, where $p$ is odd. To show that such configurations
of branes are possible, we have to slightly generalise the rules
for D-strings to end on Dp-branes \cite{aehw97} to allow for coincident
 higher-dimensional branes. 

The derivation follows \cite{str95} and is based on the fact that RR 
two-form $C$ couples to odd-dimensional 
branes via world-volume Chern-Simons term \cite{dou95}:
\be
I=\int_{D1}C\wedge\tr{}{\bf 1}+\frac{1}{2\pi}\int_{D3}C\wedge\tr F
+\frac{1}{8\pi^2}\int_{D5}C\wedge\tr F\wedge F+\ldots,
\ee
where $F$ is the field strength of the internal gauge field. 
Consequently, the total two-form RR charge density is equal to
\be
J^{i0}(y)=k\int_{D1}dx^i\,\D(y-x)+\frac{1}{2\pi}\int_{D3} 
dx^i\wedge\tr F\,
\D(y-x)+\frac{1}{8\pi^2}\int_{D5} 
dx^i\wedge\tr F\wedge F\,\D(y-x)+\ldots,
\ee
where $k$ is the RR charge of the D-string.
If the D-string has an end point on a higher dimensional D-brane, 
the charge conservation condition,
\be
\d_iJ^{i0}=0,
\ee
can only be satisfied if the field strength $F$ has a singularity
at the point where the D-string terminates. Integrating over an 
eight-sphere surrounding the end point of the D-string, we get:
\be
k=\left\{
\begin{array}{ll}
-\frac{1}{2\pi}\int\tr F,&~~~\mbox{if the D-string terminates on a D3-brane}
\\-\frac{1}{8\pi^2}\int\tr F\wedge F,
&~~~\mbox{if the D-string terminates on a D5-brane}\\
\ldots &
\end{array}
\right. ,
\ee
where the integral is over the sphere surrounding the end-point of the 
D-string on the Dp-brane. The right hand side is the monopole charge,
so, the end point of a D-string looks like a magnetic monopole.

The above consideration concerns D-strings ending on one stack of
Dp-branes. The D-string stretched between two stacks
of $N_1$ and $N_2$ coincident branes describes a monopole of $U(N_1)$,
which corresponds to one end of the string, and an anti-monopole of
$U(N_2)$, which corresponds to the other end. The monopole
arising due to symmetry breaking $U(N_1+N_2)\rightarrow U(N_1)
\times U(N_2)$ has precisely the same quantum numbers.  

In the subsequent sections we classify all possible defects, both
monopoles and instantons, on Dp-branes with $p\leq 6$.

\newsection{2-branes}

The only possible defects on D2-branes are instantons  
classified by $\pi_1(U(N))=Z$.
Obviously, they are associated with $U(1)$ factor in $U(N)$.
Relevant degrees of freedom are described at low energies by
free super-symmetric Abelian gauge theory. The localised
classical field configurations carrying the instanton charge are
unstable in the free theory. 
This instability in the D-brane picture is reflected in the fact
that the fundamental string has a tachyonic mode for a 0-brane embedded 
in the world-volume of a 2-brane \cite{pol96}.

It is worth
mentioning that D0-branes embedded in Dp-branes can also be
associated with delocalized magnetic flux   
\cite{tow95,flux1,flux2,bac98,lt99} and a 2-brane carrying a 0-brane charge 
can be described in this way \cite{tow95}. 
The configurations of this type either require the finiteness
of the D-brane volume or carry an infinite RR charge and do not
fall into the class of topological defects that we consider.

\newsection{3-branes}

The only non-trivial $\pi_2$ group is $\pi_2(G/H)=\pi_1(H)/\pi_1(G)
=Z^{m-1}$, which is responsible for magnetic charges
in ${\cal N}=4$, $d=4$ Yang-Mills theory. Different states carrying
these charges were studied
from the string-theory point of view in great detail 
\cite{dia96,sch96,ber97,hhs98,ko98,bk98,bhlms99}.

\newsection{4-branes}

The D4-branes possess internal instanton charges associated with
$\pi_3(U(N))=Z$. It worth noting that
we regard instantons as static field configurations with finite
energy, rather than Euclidean solutions with finite action in one
dimension lower. These two types of topological defects are related by
T-duality in the time direction which transforms bound states of
D0 and Dp branes into bound states of D-instantons and D(p-1) branes,
configurations recently discussed in the context of AdS/CFT
correspondence \cite{bg98,chw98,kl98,ps98,bgkr98,dkmv98,bc98,akh98,lt99}.

If there are coincident branes (some $N_i\geq 2$), the instantons can
always be colour-rotated to the unbroken subgroup and there are stable
classical solutions with any instanton charge. The homotopy group
$\pi_3(G/H)$ is trivial in this case. 
The reasoning is the same as in Appendix~A.
If all the branes are separated  ($N_i=1$) we have the short exact 
sequence
\be\label{es1}
\begin{array}{ccccccc}
\pi_{p-1}(U(1)^N) & \la & \pi_{p-1}(U(N)) & \stackrel{f_1}{\la} &
\pi_{p-1}(U(N)/U(1)^N) & \la &
\pi_{p-2}(U(1)^N)\\
\| &&   &&  && \| \\
0  &&   &&  && 0
\end{array}~~~(p>3).
\ee
Thus, $\pi_3(U(N)/U(1)^N)=\pi_3(U(N))$ and corresponding topological
defects are instantons of the broken group. 

Instantons in the Higgs model, 
as classical solutions of the field 
equations, are unstable and tend to shrink to a point.
The size of an instanton and other instanton
moduli are described by vacuum expectation values of fields 
on the 0-brane world line coming
from fundamental strings stretched between the 0- and 4-branes 
\cite{dou96,bp97,akh98}. The reason for the instability of an instanton
on separated branes is the generation of masses for moduli fields 
(except for the instanton coordinates) 
proportional to the separation between the 0- and 4-branes \cite{bp97,akh98}.
The existence of these masses is what 
forces the instanton to shrink to a point. 

\newsection{5-branes}\label{u2}

It is necessary to consider the cases $N=2$ and $N>2$ separately.

\subsection{$N>2$}

If $N>2$, the homotopy groups $\pi_4(G/H)$ can be determined from the
long exact sequence
\be\label{es12}
\begin{array}{ccccccccc}
\pi_{4}(G) & \la & \pi_{4}(G/H) & \stackrel{f_2}{\la} &
\pi_{3}(H) & \la &
\pi_{3}(G) & \la & \pi_3(G/H)\\
\| &&&& \|  && \| && \| \\
0  &&&& Z^{n_2}    && Z && 0
\end{array}~~~(N>2, n_2>0),
\ee
where $n_2$ is the number of branes with RR charge $N_i\geq 2$.
We assume that $n_2>0$, otherwise $\pi_4(G/H)=0$, as follows from the
same exact sequence.
We find that 
$$
\pi_4(G/H)=\pi_3(H)/\pi_3(G)=Z^{n_2-1}~~~.
$$ 
Therefore, the  homotopy groups are
responsible for monopoles. The states carrying monopole charges are
described by D-strings stretched between adjacent stacks of coincident
branes. 

\subsection{$N=2$}

The homotopy groups $\pi_4(U(2))=\pi_4(U(2)/U(1)^2)=Z_2$
(the second equality follows from the short exact sequence \rf{es1})
allow for $Z_2$ instantons on 5-branes\footnote{Discrete theta-angle
in 5-dimensional field theory on a 4-brane associated with $\pi_4(U(2))$
was discussed in \cite{dkv96}.}.
In the field theory language,
the non-trivial element of $\pi_4(U(2))$ can be represented as
an instanton-anti-instanton pair rotated through $2\pi$ and then 
annihilated \cite{nah80}. 

As far as $\pi_4$ is considered, we can make no difference between
$U(2)$ and $SU(2)=S^3$.
The topological classification of maps from $S^4$ to $S^3$ is given
by Freudenthal theorem \cite{math}.
The ``easy part'' of the Freudenthal theorem states that the suspension 
map
\be
\Sigma: \pi_{3}(S^2)\la\pi_4(S^3)
\ee
is an epimorphism. It follows from the ''difficult part`` of the 
Freudenthal theorem that
the kernel of this epimorphism is
the mapping with the instanton number $2$, which
infers $\pi_4(S^3)=Z_2$.
The suspension of the 
mapping with the instanton number $1$ represents 
the only non-trivial element of $\pi_4(S^3)$.

To construct the non-trivial map from $S^4$ to $S^3$ we can use the
isomorphism $\pi_3(S^3)=\pi_3(S^2)$. If $U(x)$ represents the canonical
generator of $\pi_3(SU(2))$, 
then $\alpha: x\lp U^{-1}\sigma^3U$ defines the
mapping from $S^3$ to $S^2$, the suspension of which, $\Sigma\alpha$,
is the unique non-trivial element of $\pi_4(S^3)$. Explicitly, the
suspension of $\alpha$ can be realized as a mapping
$\Sigma\alpha: (x,\tau)\lp U^{-1}\e^{2\pi i\sigma^3\tau}U$.

If $\tau$ is thought of as one of the coordinates on the 5-brane, the 
gauge transformation at infinity which is equal to unity for
$\tau<0$ and for $\tau>1$ and to 
$V=U^{-1}\e^{2\pi i\sigma^3\tau}U$ for
$0<\tau<1$ defines non-trivial element of the homotopy group 
$\pi_4(U(2))$. 
The field configuration which  goes to the pure 
gauge
$$
\Phi\la  V^{-1}\phi V,~~~~A\la V^{-1}d V
$$
at infinity will carry $Z_2$ charge. Here
$\phi=L\sigma^3$, where $L$ is the distance between branes.
For a given $\tau$ the fields in this configuration 
are that of an
instanton-anti-instanton pair with the instanton rotated 
through the angle $2\pi\tau$, since the fields of the rotated 
instanton define the pure gauge 
at infinity $\e^{2\pi i\sigma^3\tau}U$, while  
the anti-instanton corresponds to the pure gauge $U^{-1}$. Thus, $Z_2$
instanton on two 5-branes can be constructed as follows. At some
point inside 5-brane volume the instanton-anti-instanton pair is
created, then instanton and anti-instanton are separated and the 
instanton is adiabatically rotated through the angle $2\pi$, after
\begin{figure}[t]
\hspace*{5cm}
\epsfxsize=7cm
\epsfbox{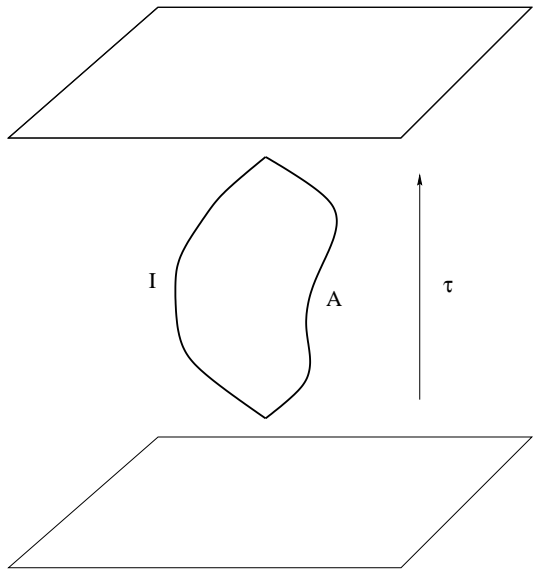}
\caption[x]{Instanton-anti-instanton pair in the 5-brane volume.}
\label{IA}
\end{figure}
which the pair is annihilated (fig.~\ref{IA}).     

This interpretation readily suggests a brane picture of the 
$Z_2$ instanton. The trajectory of an instanton corresponds
to the D1 brane, as can be established using T-duality. 
The gauge orientation of the
instanton is described by a vacuum expectation value of one of 
the massless modes of the 1-5 fundamental strings.
Thus, we conclude that the $Z_2$ instanton is a closed D1 brane
inside two coincident D5 branes (or between two separated D5 branes) with
a non-trivial configuration of low-energy modes of the fundamental string
stretched between the D1 and D5 branes. It is these modes of the 
fundamental string
that are responsible for the topological stability of the $Z_2$ instanton.
The topological stability does not imply that there is a non-singular
field configuration with minimal energy in the sector with non-trivial
$Z_2$ charge. 
The D-string loop, most probably, will tend to shrink to a
point, although it will not be able to annihilate.
\label{cpn}

\newsection{6-branes}

As follows from \rf{bott}, \rf{es1}, $\pi_5(U(N))=Z=\pi_5(U(N)/U(1)^N)$,
except for $N=2$ when $\pi_5(U(2))=Z_2=\pi_5(U(2)/U(1)^2)$. These
homotopy groups are responsible for instantons ($Z_2$-instantons in
$N=2$ case) on D6 branes. Scaling arguments show that localised 
instantons are unstable \cite{tay97}. On the string side, this is
reflected in the fact that D0 branes always repel from D6 branes
\cite{pol96}. Again, as in the case of instantons on D2 branes, it
is possible to construct delocalized stable solutions of low-energy
equations of motion \cite{tay97}.

Considering the homotopy groups $\pi_5(G/H)$ we have to distinguish 
two cases: $n_3>0$ and $n_3=0$. Here $n_3$ denotes a number of D6 branes 
with RR charge $N_i\geq 3$. 

\subsection{$n_3>0$}

Consider the exact sequence
\be\label{esD6}
\begin{array}{ccccccccc}
\pi_{5}(H) & \la & \pi_{5}(G) & \stackrel{f_1}{\la} &
\pi_{5}(G/H) & \stackrel{f_2}\la &
\pi_{4}(H) & \la & \pi_4(G)\\
\| && \| &&&& \| && \| \\
Z^{n_3}\otimes(Z_2)^{\nu_2}  && Z  &&&& (Z_2)^{\nu_2} && 0
\end{array}~~~(n_3>0),
\ee
where $\nu_2$ is the number of branes with RR charge $N_i=2$. Since
the generator of the homotopy group $\pi_5(U(N))$ can be represented by
a $U(N)$ matrix of the form
$$
U(x)=\left(
\begin{array}{cc}
\Omega(x) & 0\\
0 & {\bf 1}
\end{array}
\right),
$$
where $\Omega$ is a $3\times 3$ matrix, the first homomorphism in the chain
\rf{esD6} covers the entire $\pi_5(G)$ group. 
Hence, $f_1$ is a trivial mapping:
$\Im f_1=0$, and the boundary homomorphism $f_2$ 
is an isomorphism in this case: 
$$
\pi_5(G/H)=\pi_4(H)=(Z_2)^{\nu_2}.
$$
Thus, there are $\nu_2$ species of $Z_2$ monopoles on such configurations
of D6 branes.

We should impose unitary gauge in order to interpret $Z_2$ monopoles in
the D-brane picture. For simplicity we consider the case of $\nu_2=1$.
Passing to the unitary gauge leaves a Dirac string with the pure gauge of
the unbroken $U(2)$ subgroup around the string
representing the non-trivial element of
$\pi_4(U(2))$. Just as in Sec.~\ref{u2}, we can use the suspension map
$\Sigma:\pi_3(U(2)/U(1)^2)\rightarrow\pi_4(U(2))$ to construct a field
configuration carrying the monopole charge. Consider the volume of a 6-brane
as the  `world-volume' of a 5-brane. The magnetic charges of 5-brane monopoles
take values in $\pi_3(U(2))$ and, hence, have a colour $U(2)$ orientation.
If we take a monopole and an anti-monopole created at some point on the
6-brane, separate them, adiabatically rotate the colour orientation
of the monopole through $2\pi$, and then annihilate the pair; the $U(2)$ 
pure gauge on the surface (homeomorphic to $S^4$) surrounding the
trajectories of the Dirac strings of the monopole and the anti-monopole
will be $U^{-1}\e^{2\pi i\sigma^3\tau}U$, where $U$ is initial colour
orientation of the monopole and $\tau$ is the `time' direction on the
6-brane. This pure gauge represents a non-trivial 
element of $\pi_4(U(2))$ (see Sec.~\ref{u2}).

Since monopoles on 5-branes are represented by open D-strings, the 
monopole loop on the 6-brane, which carries $Z_2$ magnetic charge,
corresponds to a cylindrical open D2 brane stretched between D6 branes
with RR charges $N_i=2$ and $N_j>2$ (such a D6 brane exists, since $n_3>0$),    
\begin{figure}[t]
\hspace*{5cm}
\epsfxsize=7cm
\epsfbox{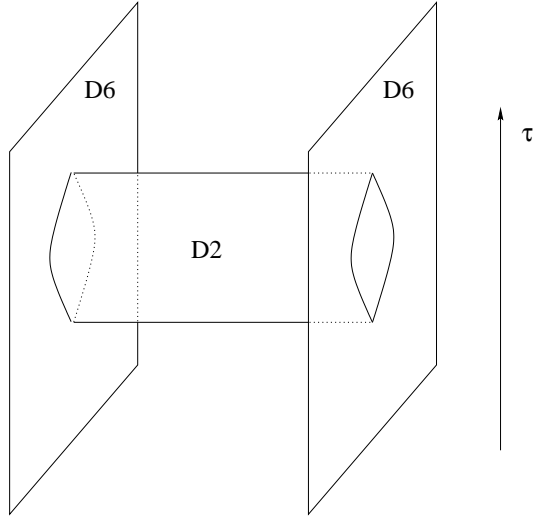}
\caption[x]{D-brane configuration describing $Z_2$ monopole on D6 brane.}
\label{ma}
\end{figure}
see fig.~\ref{ma}. The D2 brane is protected from shrinking to a point
by vacuum expectation values of massless modes of 2-6 fundamental strings.

\subsection{$n_3=0$}

In this case $G=U(2\nu_2+\nu_1)$ and $H=U(2)^{\nu_2}\times U(1)^{\nu_1}$,
where $\nu_1$ is a number of charge 1 branes. We can assume that $\nu_2>0$,
since the case of $U(N)/U(1)^N$ is considered above.

The first homomorphism in the exact sequence \rf{esD6} is trivial now,
since $n_3=0$, and we conclude that $\pi_5(G/H)/Z=(Z_2)^{\nu_2}$. This
leaves two possibilities: (1) $\pi_5(G/H)=Z\otimes (Z_2)^{\nu_2}$
or (2) $\pi_5(G/H)=Z\otimes (Z_2)^{\nu_2-1}$. In a 
special case of $\nu_1=\nu_2=1$ the vacuum manifold is 
$U(3)/U(2)\times U(1)={\bf CP}^{2}$ and the latter possibility is
realized: $\pi_5({\bf CP}^{2})=Z$ \cite{dnf}.. In general, we also have:
\be\label{pi5}
\pi_5(G/H)=Z\otimes (Z_2)^{\nu_2-1}.
\ee
The proof of this equality is sketched in Appendix~B.

Part of the topological defects carrying charges of homotopy groups
\rf{pi5} are instantons and are described by embedded D0 branes and
part of them are $Z_2$ monopoles described by open D2 branes stretched between
adjacent charge 2 D6 branes. 

There is, however, one subtlety. Consider,
for brevity, the case of $\nu_2=1$, so that $G=U(N)$ and $H=U(2)\times
U(1)^{N-2}$. The homotopy group $\pi_5(U(N)/U(2)\times U(1)^{N-2})$ 
is $Z$, so apparently there are no monopoles. But, since 
$\pi_5(U(N)/U(2)\times U(1)^{N-2})/\pi_5(U(N))=\pi_4(U(2))=Z_2$, the
group $\pi_5(U(N))$ forms $2Z$ subgroup of 
$\pi_5(U(N)/U(2)\times U(1)^{N-2})$. Therefore, only defects with even
charge are related to large $U(N)$ gauge transformations at infinity
and can be interpreted as instantons of the broken symmetry group.
The instanton number, $k=\frac{1}{48\pi^3}\int\tr F\wedge F\wedge F$,
is equal to one half of the topological charge, if we normalise it 
in such a way that the generator of $\pi_5(U(N)/U(2)\times U(1)^{N-2})$ 
has a unit charge. 
So, $k$ D0 branes embedded into D6 brane carry charge $2k$. 
Odd-charged defects cannot be smoothly transformed into the unitary
gauge. A Dirac string with $Z_2$ flux is always left. In a sense,
odd-charged defects can be interpreted as fractionally charged
instantons. 

\newsection{Discussion.}

\begin{table}[t]
\begin{center}
\begin{tabular}{|c|c|c|c|} \hline
Dp brane & & Type of & \\
configuration & \raisebox{1.5ex}[0pt]{Homotopy group} & topological defects &
\raisebox{1.5ex}[0pt]{D-brane description} \\ \hline \hline
${\bf p=2}$ & $\pi_1(U(N))=Z$ & $U(1)$ vortices & D0 branes \\ \hline \hline 
${\bf p=3}$ & $\pi_2(G/H)=Z^{m-1}$ & monopoles & open D1 branes \\ 
\hline\hline
${\bf p=4}$ & $\pi_3(U(N))=Z$ & & \\ 
$N\geq 2$ & $\pi_3(U(N)/U(1)^N)=Z$ & \raisebox{1.5ex}[0pt]{instantons} &
\raisebox{1.5ex}[0pt]{D0 branes}  \\ \hline\hline
${\bf p=5}$ & & & \\
 $N>2$ & {$\pi_4(G/H)=Z^{n_2-1}$} & {monopoles} & {open 
D1 branes} \\
 $n_2>0$ & & & \\ \hline
${\bf p=5}$ & $\pi_4(U(2))=Z_2$ & & \\
 $N=2$ & $\pi_4(U(2)/U(1)^2)=Z_2$ &
 \raisebox{1.5ex}[0pt]{$Z_2$ instantons} & 
 \raisebox{1.5ex}[0pt]{closed D1 branes} \\ \hline\hline
${\bf p=6}$ & $\pi_5(U(N))=Z$ & & \\
 $N\geq 3$ & $\pi_5(U(N)/U(1)^N)=Z$ & \raisebox{1.5ex}[0pt]{instantons} &
\raisebox{1.5ex}[0pt]{D0 branes} \\ \hline
${\bf p=6}$ & $\pi_5(U(2))=Z_2$ & & \\
 $N=2$ & $\pi_5(U(2)/U(1)^2)=Z_2$ & \raisebox{1.5ex}[0pt]{$Z_2$ instantons} &
 \raisebox{1.5ex}[0pt]{?} \\ \hline 
${\bf p=6}$ & & & \\
 $n_3>0$ & \raisebox{1.5ex}[0pt]{$\pi_5(G/H)=(Z_2)^{\nu_2}$} & 
 \raisebox{1.5ex}[0pt]{$Z_2$ monopoles} &
 \raisebox{1.5ex}[0pt]{cylindrical D2 branes} \\ \hline
 & & $Z_2$ monopoles & cylindrical D2 branes \\ \cline{3-4}
${\bf p=6}$ & & instantons & \\
 $n_3=0$ & $\pi_5(G/H)=Z\otimes (Z_2)^{\nu_2-1}$ &
($Z$ charge even) & \raisebox{1.5ex}[0pt]{D0 branes} \\ \cline{3-4}
$\nu_2>0$ & & $Z_2$ monopoles & \\
 & & ($Z$ charge odd) & \raisebox{1.5ex}[0pt]{?} \\ \hline
\end{tabular}
\end{center}
\caption{Topological classification of soliton states
on D-branes; $n_k$ is a number of Dp branes with RR charge $N_i\geq k$,
$\nu_k$ is a number of Dp branes with RR charge $N_i=k$.}
\end{table}

We have presented a complete classification of topological charges on stable
BPS D-branes of type II string theories. In almost all
cases it is possible to construct D-brane configuration representing
a state with a given topological charge. The results are summarised
in Table~1. 

Generically, embedded topological defects are described either by
D0 branes in type IIA theory or by open D1 branes in type IIB
theory and there are BPS states in any charged sector preserving some
part of the super-symmetry. 

The exceptions to this rule are $Z_2$ instantons on
D5 branes and $Z_2$ monopoles on D6 branes, which are described
by closed D1 branes and by cylindrical open D2 branes, respectively.
These branes are necessarily curved
and therefore cannot preserve the
super-symmetry. The non-trivial topology of these brane configurations 
resides in the massless modes of the fundamental strings stretched 
between the embedded and embracing branes.

We also find topological defects on certain configurations of D6 branes
which can be interpreted as fractionally charged instantons and, on
the other hand, share common features with $Z_2$ monopoles.
We have not been able to find a D-brane interpretation of these objects.

\subsection*{Acknowledgements}

We are grateful to J.~Abmj{\o}rn for hospitality at The Niels Bohr Institute
where this work has been completed.  We also thank Sandy Rutherford for 
a helpful discussion.

\setcounter{section}{0}
\appendix{Stable homotopy groups of flag manifolds}
\label{a}

In this Appendix, we shall sketch the calculation of the homotopy
groups $\pi_n(G/H)$, $G=U(N)$, $H=U(N_1)\times \ldots\times U(N_m)$ 
for $n\leq 5$,
$N_i>2$. These groups can be calculated from
the exact sequence
of the fibre bundle $$G\stackrel{H}{\la}G/H~~~.$$

For odd $n=2k+1$ we consider the exact sequence
\begin{equation}
\begin{array}{ccccccc}
\pi_{2k+1}(H) & 
\stackrel{f_0}{\la} & \pi_{2k+1}(G) &
\stackrel{f_1}{\la} & \pi_{2k+1}(G/H) & 
\stackrel{f_2}{\la} & \pi_{2k}(H) \\
\| && \| &&&& \|   \\
Z^m  && Z &&&& 0
\end{array}.
\end{equation}
First, we note that the image of any of the $m$ generators of $\pi_{2k+1}(H)$
under the homomorphism $f_0$ is the canonical generator of $\pi_{2k+1}(G)$. 
Indeed, the upper left corner immersion $U(N_1)\rightarrow U(N)$ 
induces the isomorphism of the homotopy groups 
$$\pi_{2k+1}(U(N_1))=\pi_{2k+1}(U(N))$$ for sufficiently large $N_1,N$. 
The above statement means that the image of the mapping $f_0$ is the
whole group $\pi_{2k+1}(G)$. The exactness of the sequence, {\it i.e.} that
$\Im f_0=\Ker f_1$, then implies that
$f_1$ is a trivial mapping, $\Im f_1=0$. In turn, $\Im f_1=\Ker f_2$, 
so $\Ker f_2=0$ and $f_2$ is an isomorphism. This is only possible if
\be
\pi_{2k+1}(G/H)=0.
\ee
which is our first result.

In the case of even $n=2k$ we consider the long exact sequence
$$
\begin{array}{ccccccccc}
\pi_{2k}(G) & \la & \pi_{2k}(G/H) & \stackrel{f_2}{\la} &
\pi_{2k-1}(H) & \la &
\pi_{2k-1}(G) & \la & \pi_{2k-1}(G/H)\\
\| &&&& \| && \| && \| \\
0  &&&& Z^m && Z && 0
\end{array},
$$
which implies that $$Z^m/\pi_{2k}(G/H)=Z~~~.$$ Consequently,
\be
\pi_{2k}(G/H)=Z^{m-1}.
\ee

\appendix{Homotopy groups $\pi_5(U(2\nu_2+\nu_1)/U(2)^{\nu_2}
\times U(1)^{\nu_1})$}
\label{b}

Consider the 
fibre bundle 
$$U(N)/U(2)^{\nu_2}\stackrel{U(1)^{\nu_1}}{\la} 
U(N)/U(2)^{\nu_2}\times U(1)^{\nu_1}~~~~,$$
and the short exact sequence 
\be\label{esb1}
\begin{array}{ccccccc}
\pi_{5}(U(1)^{\nu_1}) & \la & \pi_{5}(U(N)/U(2)^{\nu_2}) & 
\la & \pi_{5}(U(N)/U(2)^{\nu_2}\times U(1)^{\nu_1}) & \la &
\pi_{4}(U(1)^{\nu_1}) \\
\| &&&&&& \| \\
0  &&&&&&  0
\end{array},
\ee
from which 
we infer that 
$$
\pi_{5}(U(N)/U(2)^{\nu_2}\times U(1)^{\nu_1})=
\pi_{5}(U(N)/U(2)^{\nu_2})
$$ 
and we can study the latter homotopy groups instead.

It is instructive to consider separately 
the case of $\nu_2=1$. From the exact
sequence \rf{esD6} we know that 
$$
\pi_5(U(N)/U(2))/\pi_5(U(N))
=\pi_4(U(2))
$$ 
and that we have to distinguish between two possibilities:

\noindent
(1) $\pi_5(U(N)/U(2))=\pi_5(U(N))\otimes\pi_4(U(2))=Z\otimes Z_2$

\noindent
or 

\noindent
(2) $\pi_5(U(N)/U(2))=Z$. 

\noindent
If $N=2$, the homotopy group of interest is $Z$, as
$U(3)/U(2)=S^5$, which means that $\pi_5(U(3))$ is the $2Z$
subgroup of $\pi_5(U(3)/U(2))$.
Suppose that $\pi_5(U(N)/U(2))=Z\otimes Z_2$ for some $N$, then
$\pi_5(U(N))$ is the $Z\otimes \{1\}$ subgroup of $\pi_5(U(N)/U(2))$.
The immersion of $U(3)/U(2)$ into  $U(N)/U(2)$ induces
a homomorphism of homotopy groups 
$$f: \pi_5(U(3)/U(2)) \rightarrow
\pi_5(U(N)/U(2)).$$ 
Since any map from $S^5$ to $U(N)$ can be
reduced to the $U(3)$ subgroup, we have: $f(2)=f(1,0)$, which leaves no room
for $f(1)$ in  $\pi_5(U(N)/U(2))$. Thus we conclude that  
$$\pi_5(U(N)/U(2))=Z$$ for any $N$.

Now, consider the fibre bundle 
$$U(N)/U(2) \stackrel{U(2)^{\nu_2-1}}{\la} 
U(N)/U(2)^{\nu_2}~~~.$$
Analysis of the exact sequence
\be
\begin{array}{ccccc}
 \pi_{5}(U(N)/U(2)) & 
\la & \pi_{5}(U(N)/U(2)^{\nu_2}) & \la &
\pi_{4}(U(2)^{\nu_2-1}) \\
\| && \| && \| \\
Z  && Z\otimes (Z_2)^{\nu_2}~{\rm or}~Z\otimes (Z_2)^{\nu_2-1} &&
(Z_2)^{\nu_2-1} 
\end{array},
\ee
demonstrates the validity of eq.~\rf{pi5}.

\end{document}